\documentclass[preprintnumbers,article,amsmath,amssymb,floatfix,10pt,prd,onecolumn, superscriptaddress,nofootinbib]{revtex4}
\usepackage[colorlinks=true, pdfstartview=FitV, linkcolor=blue, citecolor=red, urlcolor=magenta]{hyperref}
\usepackage{bbm}
\usepackage{amsfonts}
\usepackage{mathrsfs}
\usepackage{latexsym}
\usepackage{epsfig}
\usepackage{epstopdf}
\usepackage{epstopdf}
\usepackage{graphicx}
\usepackage{amssymb}
\usepackage{amsmath}
\usepackage{dcolumn}
\usepackage{bm}
\usepackage{color}
\usepackage{comment}
\usepackage{xcolor}
\begin{document}
\title{\bf Ricci inverse gravity wormholes}

\author{G. Mustafa}
\email{gmustafa3828@gmail.com}\affiliation{Department of Physics, Zhejiang Normal University, Jinhua 321004, People's Republic of China}

\begin{abstract}
The current study deals with the new wormhole solutions in the background of fourth order new modified Ricci inverse gravity. Two new classes of the wormhole solutions are analyzed by showing the valid region for the main part of wormhole geometry under the affect of involved parameters. The embedded diagrams for both generic shape functions are also presented, which are connecting upper and lower Universes. In order to check the existence of these wormhole solutions, energy conditions are included in the current analysis. In the maximum regions, all energy conditions are violated, which confirms the presence of exotic matter in the background of Ricci inverse gravity. Stability analysis for both wormhole solutions is explored within the scope of speed of sounds parameters. Conclusively, some highlights from this research confirm the realistic nature and viability of these wormhole solutions in Ricci inverse gravity.  
\end{abstract}

\maketitle

\date{\today}


\section{Introduction}

The term ``wormhole" (WH) was first emerged in a superbly written paper entitled ``Classical physics as geometry" by Misner and Wheeler \cite{1}. Herman Weyl \cite{2} introduced the notion of WHs in his effort to elucidate the topological nature of the electromagnetic field.  Seeking to characterize a particle within the framework of general relativity (GR) while avoiding singularities.  Einstein and Rosen postulated a conceptualization where physical space consisted of two identical sheets, and the particle was conceived as a region of space that served as a bridge between the two sheets. We call this connection the bridge \cite{3}. WHs are theoretical structures that connect two unique universes or two separated sections of the same universe by a neck region. This is explained on a wider scale by solutions of the Einstein field equations. It is predicted that such WHs are not traversable due the occurrence of singularity. The geometry of Schwarzschild WH being the non-traversable structure is described by Fuller and Wheeler \cite{4} with the help of Kruskal constituents. In order to explore the existence of WHs having traversable nature, enormous researchers considered the theoretical observations from Schwarzschild WH as well as the Einstein-Rosen bridges. Morris and Thorne \cite{Morris} introduced a WH solution in 1988 in a static, spherically symmetric shape with a traversable throat in the center. Both solutions were created within the framework of GR, but their existence necessitates the presence of exotic matter, i.e. the null energy condition (NEC) must be violated in order to achieve a stable and traversable structure. This fact called into doubt the validity of such solutions as realistic physical systems within the strict area of GR.

Based on some cosmological observations (supernova Type Ia, Cosmic microwave background radiation, large-scale structure) \cite{7,8}, it is clear that our universe is experiencing an accelerating expansion due to a mysterious form of energy dubbed as dark energy. Alternative theories of gravity are considered to be the most remarkable and successful ways to uncover the mysteries of the universe and provide evidence for its expansion. These theories are derived by adding or replacing the function of curvature invariant in the geometric part of the Einstein-Hilbert action. To examine the dark energy as well as the continuous expansion of the universe, $f(R)$ theory ($R$ is the Ricci scalar) is the simplest approach \cite{9}. Pavlovi\'{c} and Sossich \cite{s10} examined the WHs for viable $f(R)$ gravity models and derived a simple expression for the violation of weak energy condition (WEC) near the throat. Zubair et al. \cite{s6} investigated the spherical WH solutions in $f(R,T)$ gravity. For noncommutative distributions, they found that the obtained WH solutions are stable, physically viable and satisfy the WH existence criteria. Mustafa et al. \cite{s1} obtained the stable spherically symmetric WH solutions in extended teleparallel gravity using Gaussian as well as Lorentzian non-commutative geometry and showed that the NEC is violated. Capozziello et al. \cite{v13} explored traversable WH solutions with two particular shape functions for power-law $f(R)$ model. Modified theories present a strong alternative to tackle the problem of exotic matter. Many solutions of traversable WH geometries in the context of modified gravity theories have been thoroughly studied in the literature \cite{bohmer2012,lobo2009,lobo2020,harko2013,kanti2011,usmani2010,rahaman2006,rahaman2012,zubair2016,ovgun2018,mustafa2021,cap1,cap2,cap3,cap4,cap5,kuhfittig2015,kuhfittig2005,ref1}.

 Dai and Dejan \cite{refjim1}  explored the real-world possibility of WHs and proposed that Sgr A* could be a WH candidate. A sensitive astronomical test to detect whether a black hole is a WH was done by John and his collaborators \cite{refjim2}  in the orbit of a pulsar orbiting the black hole induced by a perturbing approach on the opposite side of the WH. Bambi and his coauthor \cite{refjim3} have shown that WH solutions are black hole mimics. Essentially, their analysis is based on recent considerable developments in the capability of probing the strong gravity region of black holes, which has motivated new research to evaluate whether some or all of the reported astrophysical black holes are WHs. Falco and his his collaborators \cite{refjim4,refjim5} derived the equations of motion of a test particle in a static and spherically symmetric WH spacetime under the influence of the general relativistic Poynting-Robertson effect in GR and extended version of gravity. They considered a particular black hole mimickers class of WHs, where the exotic matter is placed in a small region close to the throat. Recently, me and my collaborators \cite{refjim6,refjim7} used the observational data of the M87 galaxy and the Milky Way galaxy within the scope of dark matter halos and found the possibility of generalized WH formation in the galactic halo region.

The Ricci-inverse gravity framework \cite{Amendola} offers a fairly new class of interesting fourth-order modified gravity model. It generalizes the Einstein-Hilbert action with a function of the Ricci scalar $R$ and the anticurvature scalar $A$. It is important to remember that the anti-curvature scalar $A$ is not the Ricci scalar $R$ inverse. Amandola et al. \cite{Amendola} examined altered gravity actions that included either positive or negative anti-curvature scalar powers. The authors demonstrated that cosmic trajectories from a decelerated phase could not smoothly enter an accelerated phase, and they proved a general no-go theorem for these models. This modified theory firstly explored by me and my collaborators \cite{RI1} for spherically symmetric peacetime, where we calculated compact stars solutions within the scope of linear equation of state. Using the reduced action approach, Das et al. \cite{RI2} demonstrated that it is unable to get around the no-go theorem for Ricci-inverse gravity models. They also discussed about how their findings will affect the cosmology of the early Universe. Souzaa and Santos \cite{RI3} calculated two different axially symmetric spacetimes with causality violation, and they shown that Ricci-inverse gravity permits the existence of closed time-like curves. Tuan Q. Do \cite{RI4} presented singularity free solutions for isotropically and anisotropically inflating Universes. Recently, Jawad and his collaborator \cite{RI5} investigated the matter-antimatter asymmetry through baryogenesis in the framework of Ricci inverse gravity. Being motivated from the study regarding WH solutions and Ricci inverse gravity, we explore some new interesting aspects of WH geometry in the background of Ricci inverse gravity in the current analysis.

The current manuscript is organized as follows: Sec. (II) provides the field equations for Ricci inverse gravity for WH geometry. Embedded diagrams for two generic WHs models are also shown in same section. In Sec. (III), we provide the energy conditions and their pictorial analysis. In Sec. (IV), stability analysis is provided by exploring speed of sounds parameters around the WH throat. Finally, the last Sec. gives the discussion of the result and concluding remark.

\section{Inverse Ricci Gravity}

In this section, we shall discuss Ricci inverse gravity and explore the filed equations. The inverse tensor of $R_{ij}$ is defined as
\begin{equation} \label{14}
A^{ij}R_{c}=\delta^{i}_{c},
\end{equation}
where $A^{ij}$ is defining the anti-curvature tensor. The modified action for Ricci-inverse based gravity was provided by \cite{Amendola} and it is defined as:
\begin{equation} \label{15}
S=\int \sqrt{-g}d^4x(\alpha A+R).
\end{equation}
In the above equation, the expression $A$ is a trace of $A^{ij}$, which is calculated as
\begin{equation} \label{16}
A^{ij}=R^{-1}_{ij}.
\end{equation}
On differentiating Eq. (\ref{16}), one can get the following relation
\begin{equation} \label{17}
\delta A^{i \tau}=-A^{i j}\left(\delta R_{j \sigma}\right) A^{\sigma \tau} .
\end{equation}
Now, from the Eq. (\ref{15}) we have the following expressions
\begin{eqnarray}\label{18}
\delta S && =\int d^4 x\left(A \delta \sqrt{-g}+\sqrt{-g} A^{i v} \delta g_{i j}+\sqrt{-g} g_{i j} \delta A^{i j}\right), \\\label{19}
&& =\int d^4 x \sqrt{-g}\left(\frac{1}{2} A g^{i v} \delta g_{i j}+A^{i v} \delta g_{i j}+g_{i j} \delta A^{i j}\right),
\end{eqnarray}
and since
\begin{equation} \label{20}
\delta R_{\epsilon \beta}=\nabla_\rho \delta \Gamma_{\beta \epsilon}^\rho-\nabla_\beta \delta \Gamma_{\rho \epsilon}^\rho,
\end{equation}
we obtain
\begin{eqnarray}\label{21}
\delta A^{i j} && =-A^{i \epsilon}\left(\nabla_\rho \delta \Gamma_{\beta \epsilon}^\rho-\nabla_\beta \delta \Gamma_{\rho \epsilon}^\rho\right) A^{\beta j}, \\\label{22}
&& =-\frac{1}{2} A^{i \epsilon}\left(g^{\rho \lambda} \nabla_\rho\left(\nabla_\epsilon \delta g_{\beta \lambda}+\nabla_\beta \delta g_{\lambda \epsilon}-\nabla_\lambda \delta g_{\epsilon \beta}\right)-g^{\rho \lambda} \nabla_\beta\left(\nabla_\epsilon \delta g_{\rho \lambda}+\nabla_\rho \delta g_{\lambda \epsilon}-\nabla_\lambda \delta g_{\epsilon \rho}\right)\right) A^{\beta j}, \\\label{23}
&& =-\frac{1}{2} A^{i \epsilon} g^{\rho \lambda}\left(\nabla_\rho \nabla_\epsilon \delta g_{\beta \lambda}-\nabla_\rho \nabla_\lambda \delta g_{\epsilon \beta}-\nabla_\beta \nabla_\epsilon \delta g_{\rho \lambda}+\nabla_\beta \nabla_\lambda \delta g_{\epsilon \rho}+\left[\nabla_\beta, \nabla_\rho\right] \delta g_{\lambda \epsilon}\right) A^{\beta j}
\end{eqnarray}
Now, by using integration by parts technique, we get
\begin{eqnarray}\label{24}
g_{i j} \delta A^{i j} && =-\frac{1}{2} g_{i j} g^{\rho \lambda}\left(\delta g_{\beta \lambda} \nabla_\epsilon \nabla_\rho\left(A^{i \epsilon} A^{\beta v}\right)-\delta g_{\epsilon \beta} \nabla_\lambda \nabla_\rho\left(A^{i \epsilon} A^{\beta v}\right)-\delta g_{\rho \lambda} \nabla_\epsilon \nabla_\beta\left(A^{i \epsilon} A^{\beta v}\right)\right.\nonumber\\&&\left.+\delta g_{\epsilon \rho} \nabla_\lambda \nabla_\beta\left(A^{i \epsilon} A^{\beta v}\right)\right) \\\label{25}
&& =\frac{1}{2} \delta g_{\iota \kappa}\left(-2 g^{\rho \iota} \nabla_\epsilon \nabla_\rho A^{i \epsilon} A_i^\kappa+\nabla^2\left(A^{i \iota} A_i^\kappa\right)+g^{\iota \kappa} \nabla_\epsilon \nabla_\beta\left(A^{i \epsilon} A_i^\beta\right)\right).
\end{eqnarray}
So finally by the variation, we have
\begin{equation} \label{26}
\delta g_{i j}\left(\frac{1}{2} A g^{i j}+A^{i j}+\frac{1}{2}\left(-2 g^{\rho i} \nabla_\epsilon \nabla_\rho A^{\sigma \epsilon} A_\sigma^j+\nabla^2\left(A^{\sigma i} A_\sigma^j\right)+g^{i j} \nabla_\epsilon \nabla_\beta\left(A^{\sigma \epsilon} A_\sigma^\beta\right)\right)\right) .
\end{equation}
On the variation of the Lagrangian
\begin{equation} \label{27}
\delta g^{i j}\left(-\frac{1}{2} R g_{i j}+R_{i j}\right)=-\delta g_{i j}\left(-\frac{1}{2} R g^{i j}+R^{i j}\right),
\end{equation}
By adopting the procedure from \cite{Amendola}, we have the following set of field equations
\begin{equation}\label{28}
R^{ij}-\frac{1}{2}Rg^{ij}-\alpha A^{ij}-\frac{1}{2}\alpha A g^{ij}+\frac{\alpha}{2}\left(2g^{\varrho i}\nabla_{\epsilon}\nabla_{\varrho}A^{\epsilon}_{\sigma}A^{b\sigma}-\nabla^2A^{i}_{\sigma}
A^{j\sigma}-g^{ij}\nabla_{\epsilon}\nabla_{\varrho}A^{\epsilon}_{\sigma}A^{\varrho\sigma}\right)={\mathcal{T}}^{ij}
\end{equation}
where $A^{\epsilon}_{\sigma}A^{b\sigma}=A^{\epsilon\tau}g_{\tau\sigma}A^{\sigma b}=A^{\epsilon\tau}A^{b}_{\tau}=A^{\epsilon\sigma}A^{b}_{\sigma}
=A^{b}_{\sigma}A^{\epsilon\sigma}$ with $8\pi G=1$.
\subsection{Inverse Ricci Gravity and WH geometry}
In order to discuss the WH geometry in the background of inverse Ricci gravity. We start with the stress tensor possessing anisotropic matter content, which is read as
\begin{equation} \label{29}
{\mathcal{T}}_{ij}=(\rho+p_{t})u_{i}u_{j}-p_{t}g_{ij}+(p_{r}-p_{t})v_{i}v_{j}.
\end{equation}
Here, $u_{i}=e^{\frac{b}{2}}\delta_{i}^{0}$ and $v_{i}=e^{\frac{\lambda}{2}}\delta_{i}^{1}$. The spherically symmetric space-time is given as
\begin{equation} \label{30}
ds^{2}=e^{a(r)}dt^{2}-e^{b(r)} dr^{2}-r^{2} d\theta^{2}-r^{2}\sin^{2}\theta d\phi^{2},
\end{equation}
Now we use the spherically symmetric Morris and Thorne WH geometry in Schwarzschild coordinates \cite{Morris}, which is realized as
\begin{equation}\label{31}
e^{a(r)}=e^{2\Phi(r)}, \;\;\;\;\;\;\;\;\;\;\;\;e^{b(r)}=\left(1-\frac{X(r)}{r}\right)^{-1},
\end{equation}
where $X(r)$ is a candidate for a shape function and $\Phi(r)$ is defining the redshift function. For the WH geometry, $r$, i.e., radial coordinate varies between $r_0\leq r<\infty$,  with $r_0$ is the throat radius. The shape function $X(r)$ must obey the following criterion to secure a traversability of WH: The shape function should be satisfied the condition $X(r_0)=r_0$ at the throat i.e. at $r=r_0$, to ensure the throat condition. For $r>r_0$ we have $1-\frac{X(r)}{r}>0$. Flaring out condition: at the throat we also need to impose that $X^{\,\prime}(r_0)<1$. Asymptotically flatness condition: $\frac{X(r)}{r}\rightarrow o$ as $r\rightarrow \infty$.
\begin{figure}
\centering \epsfig{file=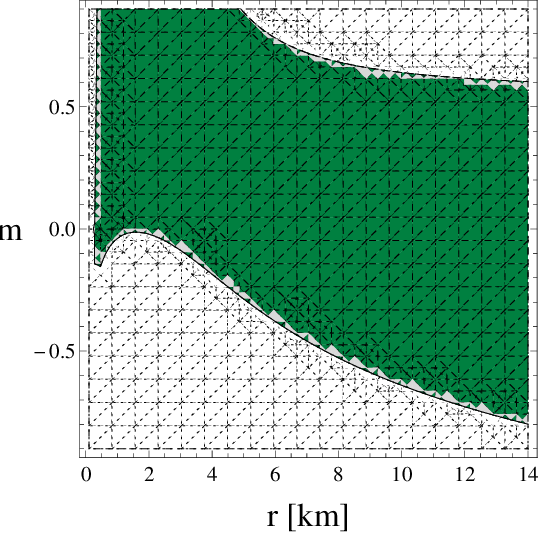, width=.48\linewidth,
height=1.8in} \epsfig{file=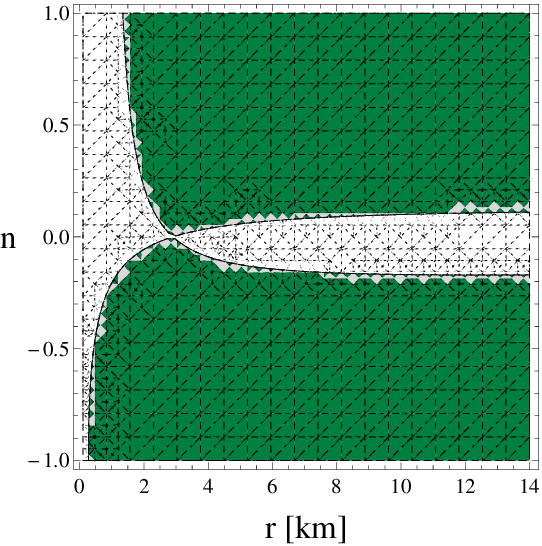, width=.48\linewidth,
height=1.8in}
\caption{\label{F1} shows the allowed-able region for both models with $\left(1-\frac{X(r)}{r}\right)>$ (shaded) and $\left(1-\frac{X(r)}{r}\right)<0$ (un-shaded).}
\end{figure}
Now, by using the material from Eq. (\ref{29}), Eq. (\ref{30}), and Eq. (\ref{31}) in Eq. (\ref{28}) on can get the following  modified field equations:
\begin{eqnarray}
\rho &=&\frac{(r-X(r)) \left(-\frac{2 \alpha  \left(\rho _1-\rho _2\right) r^3}{(r-X(r))^2}+\frac{4 \alpha  \rho _{13} r^2}{\rho _{12}}-\frac{X'(r)}{r^2-r X(r)}\right)}{r}, \label{32}\\
p_r&=&-4 \alpha  r \left(\frac{r \left(p_6 r^4-p_9 r^3 X(r)+p_5 r^2 X(r)^2-p_4 r X(r)^3+p_3 X(r)^4\right)}{p_{10}^4}+\frac{p_2}{p_1^3}\right),\label{33}\\
p_t&=&-\frac{r \left(-8 \alpha  \left(p_{12}+p_{13}\right) r^2+8 \alpha  p_{17} r^2+p_{11}\right)}{4 (r-X(r))}\label{34}.
\end{eqnarray}
where $\rho _i$, $\{i=1,...,13\}$ and $p _i$, $\{i=1,...,17\}$ are given in the Appendix. In the current study, we consider the specific redshift function to avoiding the any difficulty due the involvement of fourth order gravity, which is expressed as \cite{r32,r33,r34}
\begin{equation}\label{35}
\Phi=-\frac{\zeta }{r},
\end{equation}
where $\chi$ is constant. For the present analysis, we are considering two newly most generic series type shape functions, which are expressed as \cite{r35,r36}
\begin{eqnarray}
X(r)&=&\frac{1}{2 n^2}\left(2 \ln (r_0) (n (\mathcal{A}_0+\mathcal{A}_1)-\mathcal{A}_0 \ln
(r))-2 n \ln (r) (\mathcal{A}_0+\mathcal{A}_1)+\mathcal{A}_0 \ln(r^2)+2
n^2 (\mathcal{A}_0+\mathcal{A}_1+\mathcal{A}_2)\right.\\\nonumber&+&\left.\mathcal{A}_0 \ln(r_0^2)\right),\label{36}\\
X(r)&=&\frac{2}{2 m^2}\left( \left(\mathcal{A}_0 \left(2 n^2
\left(r^m-r_0^m\right)^2-2 m n
\left(r^m-r_0^m\right)+m^2\right)+m
\left(\mathcal{A}_1 \left(-2 n r^m+2 n r_0^m+m\right)+\mathcal{A}_2 m\right)\right)\right.\\\nonumber&+&\left.m
(\ln (r)-\ln (r_0))\left(\mathcal{A}_0 m (\ln
(r)-\ln (r_0))-2 \left(\mathcal{A}_0 \left(2 n
\left(r_0^m-r^m\right)+m\right)+\mathcal{A}_1 m\right)\right)\right),\label{37}
\end{eqnarray}

where $0.1<\mathcal{A}_0 <\mathcal{A}_1 <\mathcal{A}_2 <0.5$, $-0.9< m <0.9$, and $-0.9<n <0.9$. For the current analysis, shape function by Eq. (\ref{36}) should be treated as model-I and shape function by Eq. (\ref{37}) should be considered as model-II. Indeed, an event horizon is not allowed in the WH structural geometry. A detailed graphical analysis  regarding the properties of of above both specific generic shape function by Eq. (\ref{36}) and Eq. (\ref{37}) is already reported in \cite{r35,r36}. Here, we only show the allowed-able region for the main part of WH geometry i.e., $\left(1-\frac{X(r)}{r}\right)$ within the scope of two different involved parameters in both shape functions. The embedding surface diagram for WH geometry is now discussed using the $t = const.$ and $\theta=2\pi$ in Eq.(\ref{30}), one can get the following relation
\begin{equation} \label{38}
ds^{2}= \left(1-\frac{X(r)}{r}\right)^{-1}dr^{2}+r^{2}d\phi^{2},
\end{equation}
The Eq.(\ref{38}) can be revised into a 3-D Euclidean spacetime, which is further defined as
\begin{equation} \label{39}
ds^{2}_{\Sigma}= dg^{2}+dr^{2}+r^{2}d\phi^{2}=\bigg(1+\bigg(\frac{dg}{dr}\bigg)^{2}\bigg)dr^{2}+r^{2}d\phi^{2},
\end{equation}
On matching Eqs. (\ref{38}-\ref{39}), one can get the following relation
\begin{equation} \label{40}
\frac{dg}{dr}=\pm \left(\frac{r}{X(r)}-1\right)^{-1/2},
\end{equation}
\begin{figure}
\centering \epsfig{file=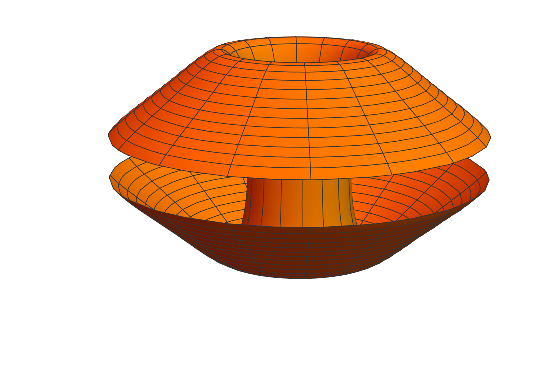, width=.48\linewidth,
height=1.8in} \epsfig{file=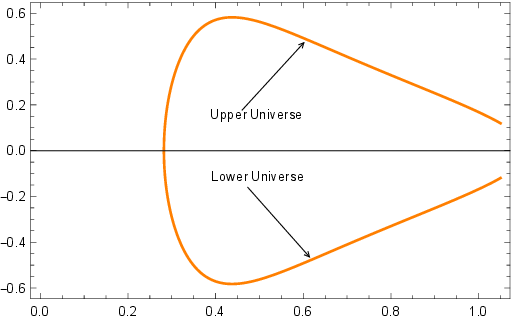, width=.48\linewidth,
height=1.8in} \caption{\label{F2} shows the embedding surface with upper and lower Universes for WH model-I.}
\end{figure}

\begin{figure}
\centering \epsfig{file=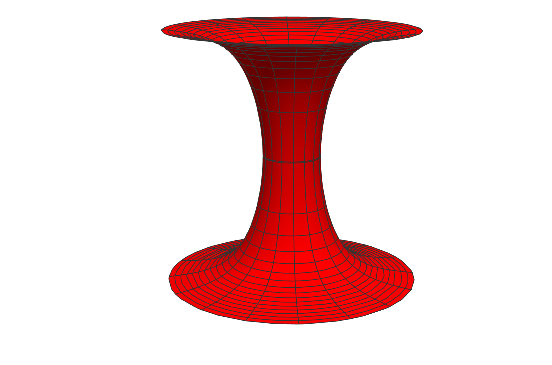, width=.48\linewidth,
height=1.8in} \epsfig{file=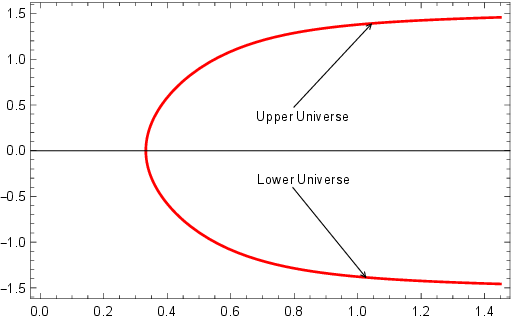, width=.48\linewidth,
height=1.8in} \caption{\label{F3} shows the embedding surface with upper and lower Universes for WH model-II}
\end{figure}
Following the discovery of the Einstein-Rosen bridge \cite{3}, Kruskal \cite{jimr3} and Szekeres \cite{jimr4} published a approach in 1960 that resembled a maximal extension of the Schwarzschild spacetime. They also published their global coordinate system, known as Kruskal coordinates, which is crucial in describing the embedding diagram.  In 1962, Fuller and Wheeler \cite{jimr5} defined the geometry of the Schwarzschild-like WH using Kruskal coordinates and showed that it is not traversable, not even by a photon. In addition, the authors included sketches of a series of WH shapes for specific collapsing, spacelike slices, showing how the WH forms and then gets ``pinched off." However, the computations for the embeddings in $R^3$ were not provided. In the present study, we create the embedded diagram of the WH using the method that was introduced by Marolf \cite{jimr7} and Peter and David \cite{jimr6}. Because of the peculiar functions involved, it is not possible to perform the analytical integration of Eq. (\ref{24}) against the computed shape functions. Using numerical techniques, we are able to show the WH embedded diagram by fixing the WH throat values for both WH models are shown in Fig. (\ref{F2}) and Fig. (\ref{F3}), respectively. \\

\begin{figure}
\centering \epsfig{file=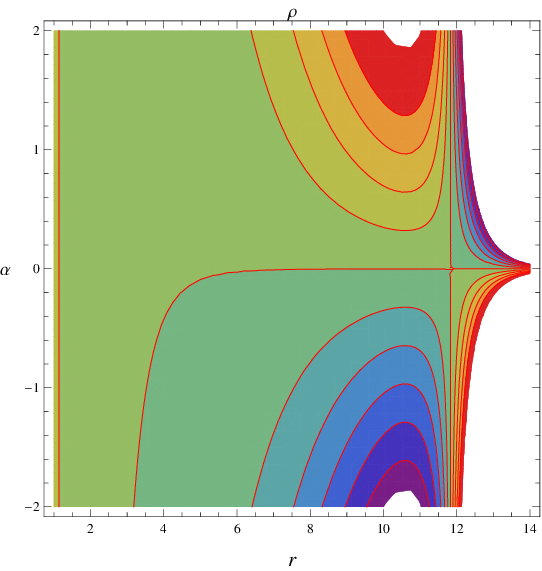, width=.4\linewidth,
height=1.8in}\epsfig{file=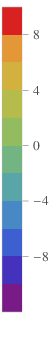, width=.08\linewidth,
height=1.8in}
\centering \epsfig{file=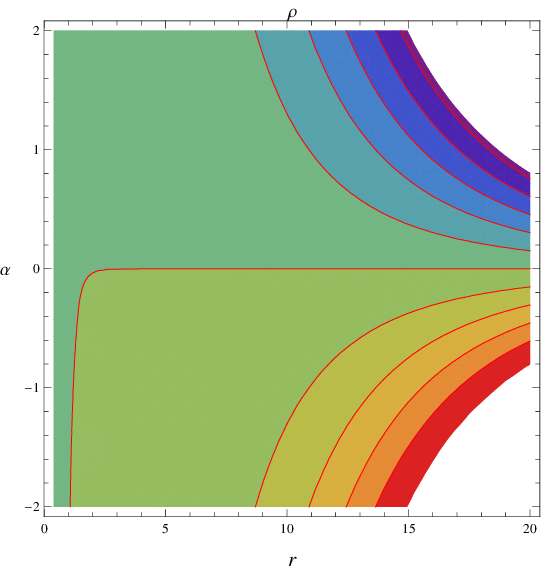, width=.4\linewidth,
height=1.8in}\epsfig{file=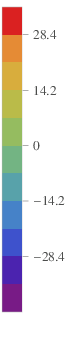, width=.08\linewidth,
height=1.8in}
\caption{\label{F4} shows the pictorial view of energy density for model-1 (left) and model-II (right).}
\end{figure}
\begin{figure}
\centering \epsfig{file=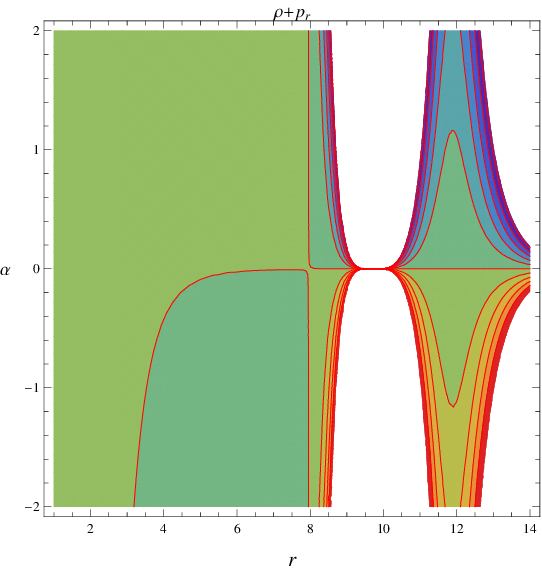, width=.4\linewidth,
height=1.8in}\epsfig{file=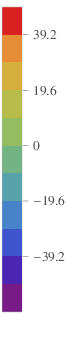, width=.08\linewidth,
height=1.8in}
\centering \epsfig{file=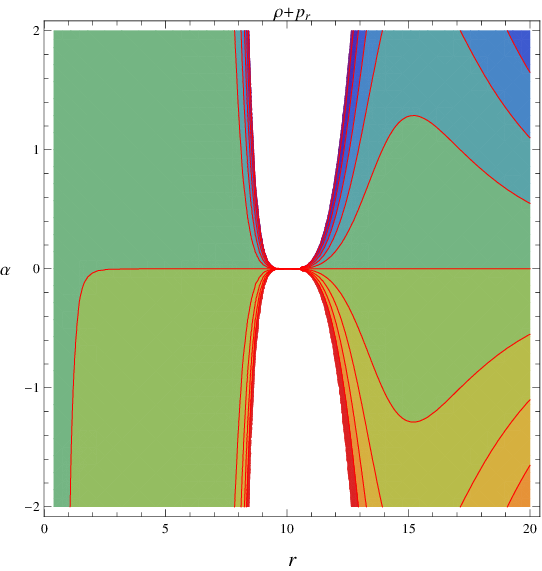, width=.4\linewidth,
height=1.8in}\epsfig{file=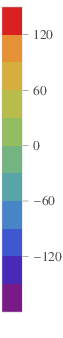, width=.08\linewidth,
height=1.8in}
\caption{\label{F5} shows the pictorial view of $\rho+p_r$ for model-1 (left) and model-II (right).}
\end{figure}
\begin{figure}
\centering \epsfig{file=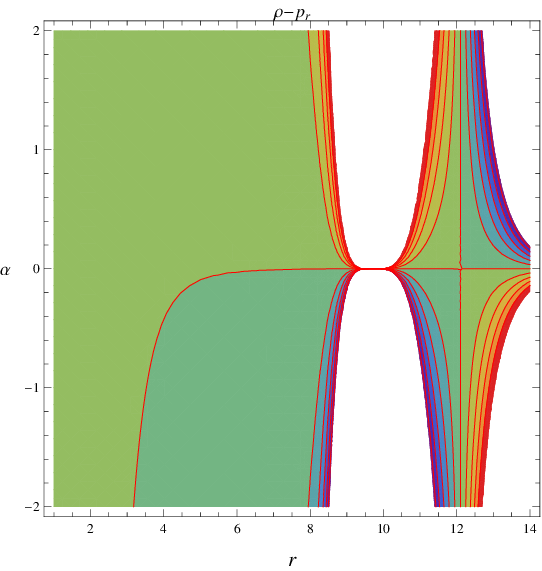, width=.4\linewidth,
height=1.8in}\epsfig{file=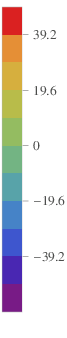, width=.08\linewidth,
height=1.8in}
\centering \epsfig{file=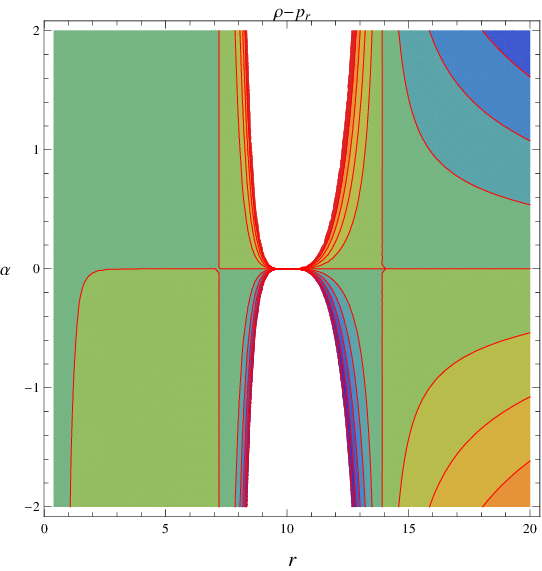, width=.4\linewidth,
height=1.8in}\epsfig{file=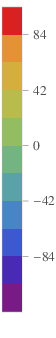, width=.08\linewidth,
height=1.8in}
\caption{\label{F6} shows the pictorial view of $\rho-p_r$ for model-1 (left) and model-II (right).}
\end{figure}

\begin{center}
\begin{table}
\caption{\label{tab1}{Summery of energy conditions for model-I and model-II.}}
\begin{tabular}{|c|c|c|c|c|c|c|c|c|}
\hline
Expressions                  & $-2\leq\alpha<0$    &$0<\alpha\leq2$  \\
\hline
    & \multicolumn{2}{|c|}{Energy conditions for Model-I} \\
\hline
 $\rho$ \;\;\;$(Fig.\ref{F4})_{left}$               & $\rho<0$                                                                                              & $\rho>0$  \\
 $\rho+p_{r}$ \;\;\;$(Fig.\ref{F5})_{left}$         & $\rho+p_{r}>0$                                                                                        & $\rho+p_{r}<0$   \\
    $\rho-p_{r}$ \;\;\;$(Fig.\ref{F6})_{left}$      & $\rho-p_{r}<0$                                                                                     &$\rho-p_{r}>0$   \\
      $\rho+p_{t}$ \;\;\;$(Fig.\ref{F7})_{left}$    & $\rho+p_{t}<0$                                                                                   &$\rho+p_{t}>0$   \\
    $\rho-p_{t}$ \;\;\;$(Fig.\ref{F8})_{left}$      & $\rho-p_{t}>0$                                                                                     &$\rho-p_{t}<0$  \\
$\rho+p_{r}+2p_{t}$ \;\;\;$(Fig.\ref{F9})_{left}$   & $\rho+p_{r}+2p_{t}<0$                                                                                  &$\rho+p_{r}+2p_{t}>0$  \\
 \hline
    & \multicolumn{2}{|c|}{Energy conditions for Model-II} \\
\hline
 $\rho$ \;\;\;$(Fig.\ref{F4})_{right}$               & $\rho>0$                                                                                              & $\rho<0$  \\
 $\rho+p_{r}$ \;\;\;$(Fig.\ref{F5})_{right}$         & $\rho+p_{r}<0$                                                                                        & $\rho+p_{r}>0$   \\
    $\rho-p_{r}$ \;\;\;$(Fig.\ref{F6})_{right}$      & $\rho-p_{r}>0$                                                                                     &$\rho-p_{r}<0$   \\
      $\rho+p_{t}$ \;\;\;$(Fig.\ref{F7})_{right}$    & $\rho+p_{t}>0$                                                                                   &$\rho+p_{t}<0$   \\
    $\rho-p_{t}$ \;\;\;$(Fig.\ref{F8})_{right}$      & $\rho-p_{t}<0$                                                                                     &$\rho-p_{t}>0$  \\
$\rho+p_{r}+2p_{t}$ \;\;\;$(Fig.\ref{F9})_{right}$   & $\rho+p_{r}+2p_{t}>0$                                                                                  &$\rho+p_{r}+2p_{t}<0$  \\
 \hline
\end{tabular}
\end{table}
\end{center}

\begin{figure}
\centering \epsfig{file=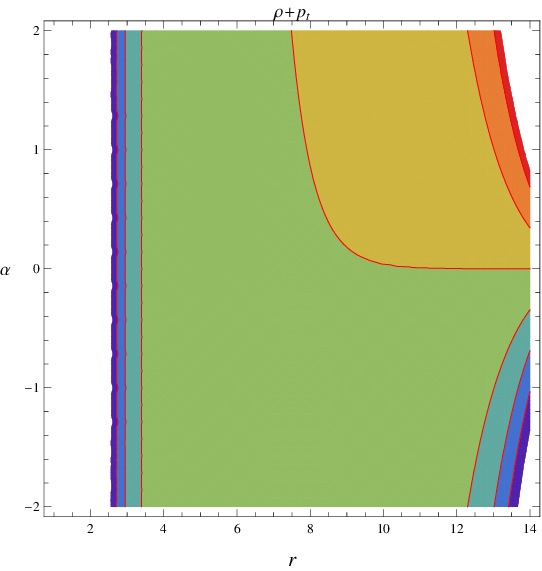, width=.4\linewidth,
height=1.8in}\epsfig{file=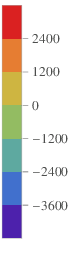, width=.08\linewidth,
height=1.8in}
\centering \epsfig{file=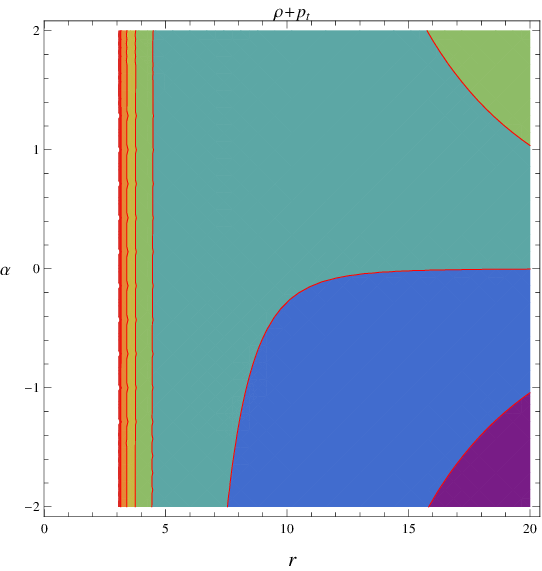, width=.4\linewidth,
height=1.8in}\epsfig{file=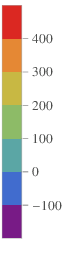, width=.08\linewidth,
height=1.8in}
\caption{\label{F7} shows the pictorial view of $\rho+p_t$ for model-1 (left) and model-II (right).}
\end{figure}
\begin{figure}
\centering \epsfig{file=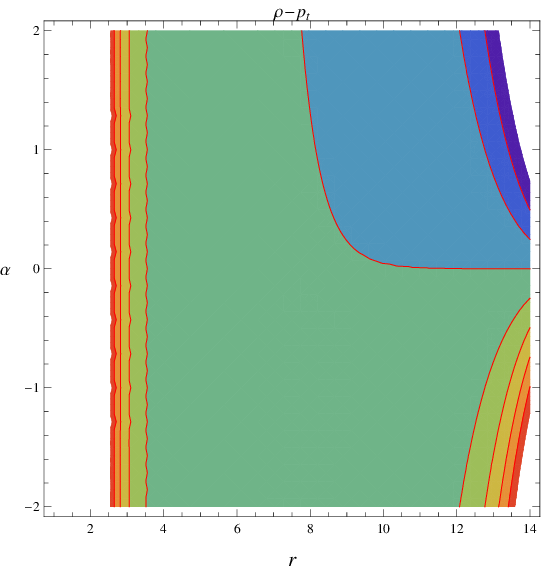, width=.4\linewidth,
height=1.8in}\epsfig{file=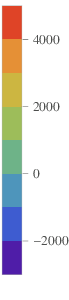, width=.08\linewidth,
height=1.8in}
\centering \epsfig{file=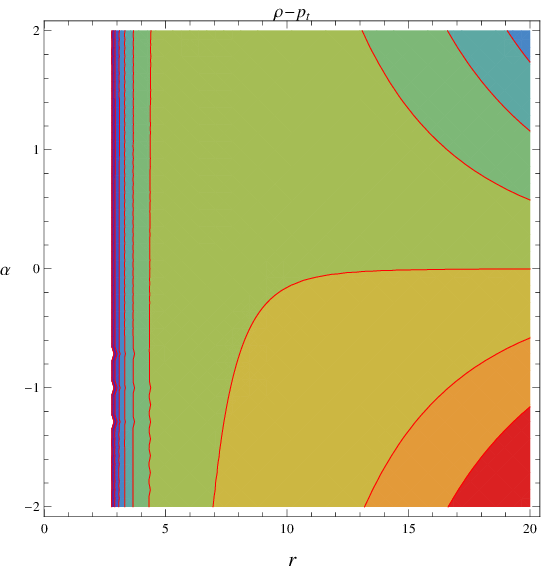, width=.4\linewidth,
height=1.8in}\epsfig{file=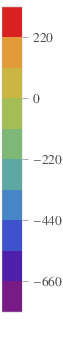, width=.08\linewidth,
height=1.8in}
\caption{\label{F8} shows the pictorial view of $\rho-p_t$ for model-1 (left) and model-II (right).}
\end{figure}
\begin{figure}
\centering \epsfig{file=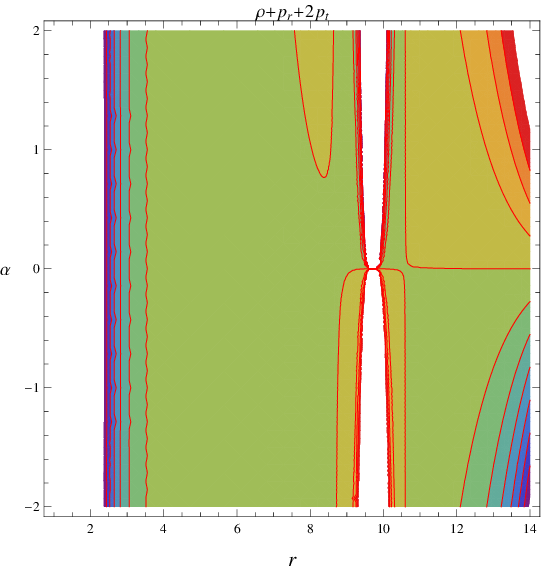, width=.4\linewidth,
height=1.8in}\epsfig{file=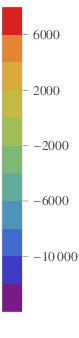, width=.08\linewidth,
height=1.8in}
\centering \epsfig{file=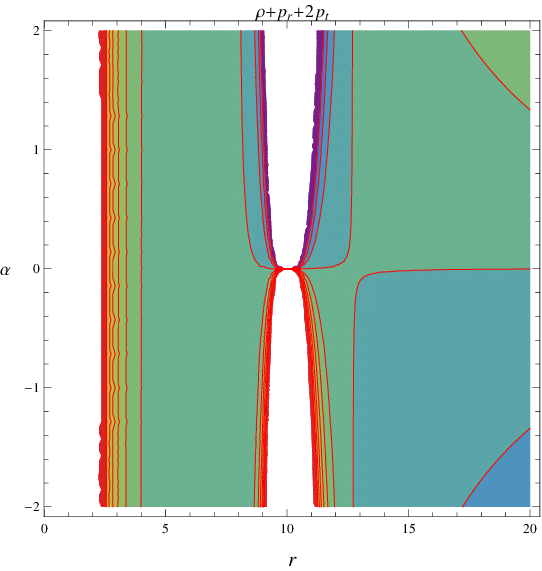, width=.4\linewidth,
height=1.8in}\epsfig{file=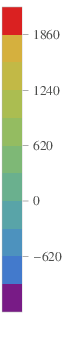, width=.08\linewidth,
height=1.8in}
\caption{\label{F9} shows the pictorial view of $\rho+p_r+2p_{t}$ for model-1 (left) and model-II (right).}
\end{figure}

\section{Pictorial analysis of Energy Conditions}

The energy conditions let us to closely examine the casual and geodesic structure of space-time, making them essential tools for inverse Ricci gravity. The Raychaudhuri equations \cite{Raychaudhuri/1955, Raychaudhuri/1955/01}, which specify the action of correspondence of gravity for timelike, spacelike, or lightlike curves, provide one way to derive such conditions. In the case of an anisotropic fluid, the following energy conditions are given for inverse Ricci gravity:
\begin{itemize}
\item   $\rho+p_k \geq0$, $\rho+\sum_k p_k\geq0$, $\forall k$ $\Rrightarrow$ Strong energy conditions (SEC).\\
\item  $\rho\geq0$, $\rho\pm p_k\geq0$, $\forall k$ $\Rrightarrow$  Dominant energy conditions (DEC) .\\
\item  $\rho\geq0$, $\rho+p_k\geq0$, $\forall k$ $\Rrightarrow$  WEC .\\
\item  $\rho+p_k\geq0$, $\forall k$ $\Rrightarrow$  NEC.\\
\end{itemize}
where $k=r,t$. The most important environment necessary for the creation of WH structures consists of the linked evolutions of the energy density and the radial and tangential pressures. For the purposes of this analysis, the evolutions of the energy conditions under the anisotropic pressure terms have been examined. The pictorial view through the related graphs for all energy conditions is given in Figs. (\ref{F4}-\ref{F9}). All these figures clearly demonstrated the regional graphical behavior for all the energy conditions. In the current analysis, due the involvement of Ricci inverse gravity the involve parameter $\alpha$ has a important role in the discussion of energy conditions. Being realized the effect of Ricci inverse gravity parameter we have plotted all the energy conditions against the specific range of $-2\leq \alpha \leq2$  with the particular values of other involved parameters in the WH solutions. The positive nature of energy density is very necessary throughout the WH configuration, which can be confirmed from the Fig. (\ref{F4}). From the Fig. (\ref{F4}), it can be seen that the energy density is positive for $0<\alpha\leq2$, while its negative for $-2\leq\alpha<0$ in the framework of model-I. For model-II, energy density is seen negative for  $0<\alpha\leq2$ and its a positive for $-2\leq\alpha<0$. It is noted from the Fig. (\ref{F5}) that the energy condition $\rho+p_r$ is observed negative, where energy density is seen positive. Further, the condition $\rho+p_r$ is observed positive, where energy density is observed negative. The violated behavior of other energy conditions is given Figs. (\ref{F6}-\ref{F8}). The SEC, i.e., $\rho+p_{r}+2p_{t}$ is also violated in some regions, which can be confirmed from Fig. (\ref{F9}). The violation of NEC and SEC confirms the presence of exotic matter, which is very necessary component for the existence of these considered WH models. All the energy conditions with negative and positive behavior for both shape functions models is also given in comprehensive way in Tab. (\ref{tab1}).
\begin{figure}
\centering \epsfig{file=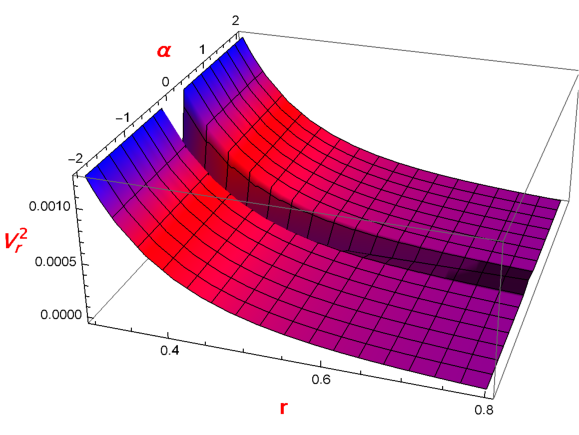, width=.48\linewidth,
height=1.8in} \epsfig{file=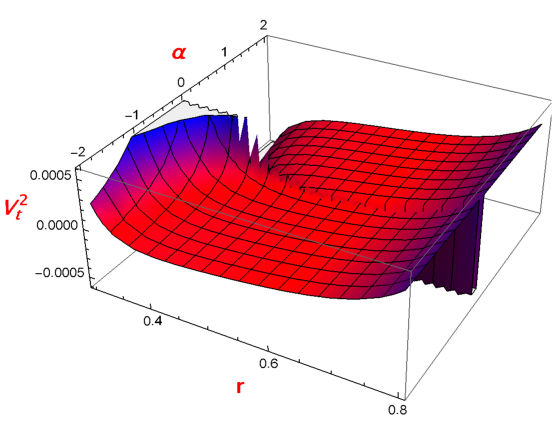, width=.48\linewidth,
height=1.8in}
\caption{\label{F10} shows the radial and tangential speed of sound parameters for model-I. }
\end{figure}
\begin{figure}
\centering \epsfig{file=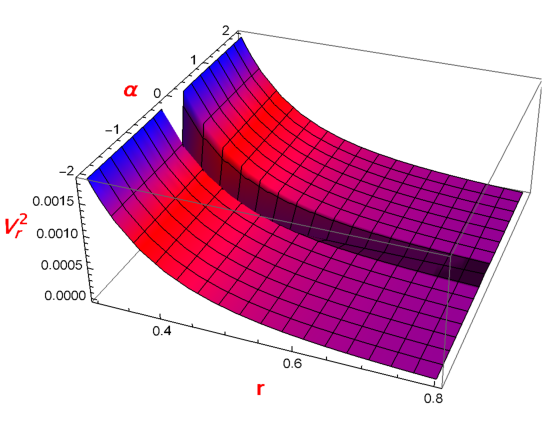, width=.48\linewidth,
height=1.8in} \epsfig{file=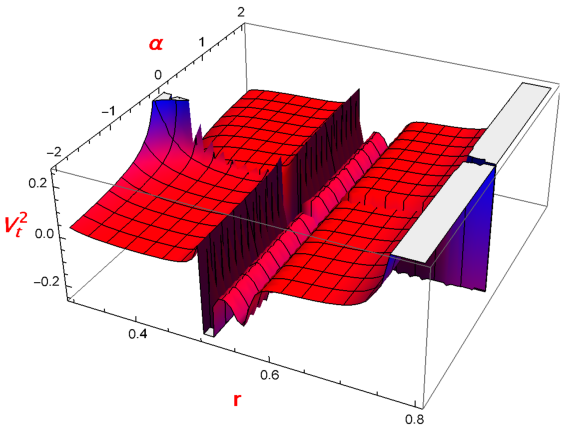, width=.48\linewidth,
height=1.8in}
\caption{\label{F11} shows the radial and tangential speed of sound parameters for model-II.}
\end{figure}
\subsection{Stability Analysis}
We examine the stability of the WH cosmic arrangement in this section. The propagation rate of sound waves via an anisotropic fluid distribution must be smaller than that of electromagnetic radiation in order for the system to be stable, i.e., $0<v_{r}^{2}<1$ and $0<v_{t}^{2}<1$, where $v_r$ and $v_t$ are calculated as \cite{plbsc}
\begin{equation}\label{42}
v_{r}^{2}=\frac{dp_r}{d\rho},\quad v_{t}^{2}=\frac{dp_t}{d\rho}.
\end{equation}
This norm is referred to as the causality condition. For the various ranges of parameter $\alpha$, the sound speed parameters $\frac{dp_r}{d\rho}$ and $\frac{dp_t}{d\rho}$ have a value between 0 and 1 around the throat. In Fig. (\ref{F10}) for model-I and Fig. (\ref{F11}) for model-II, the sound speed parameters are shown graphically. It is important to note that both the speed sound parameters and the observed WH throat radius in Fig. (\ref{F2}) and Fig. (\ref{F3}) correspond well with each others. As a result, both WH solutions provide the stable configurations.

\section{Discussion}
We have studied a recently proposed novel fourth order Ricci-inverse gravity, which introduces a very innovative geometrical component $A$ known as the anti-curvature scalar. In this letter, we have studied WH geometry under Ricci-inverse gravitational framework. In order to achieve a physically interesting consequence, we have used specific shape functions to produce WH geometry. The gravitational lensing, time dilation, exotic matter, and the displacement of matter are the instances of observational evidence for WH geometry, or theoretical indications that may suggest the existence or presence of a WH. It is necessary to mention that exotic matter has the leading role for the existence of WHs solution in the current analysis. The creation of two new generic WH solutions has released a significant influence in the field of WH research. \\

It has noted that both the considered shape functions satisfy the basic structure WH geometry according to Morris and Thorne criterion. The valid region has shown for two different involved parameters $m$ and $n$ in Fig. (\ref{F1}) for both model-I (left) and model-II (right). The shaded region for both the shape functions has allowed the existence of WH geometry, where all the conditions have meet the required conditions. Interestingly, embedded diagrams for both the considered shape functions have plotted in Fig. (\ref{F2}) and Fig. (\ref{F3}) for upper and lowered Universes. \\

All the energy conditions have been examined in Figs. (\ref{F4}-\ref{F9}). Some of the key attributes regarding energy conditions are listed as: The positive nature of energy density can be seen from the Fig. (\ref{F4}) for both models in some region against parameter $\alpha$. It has observed from the Fig. (\ref{F5}) that the energy condition $\rho+p_r$ has revealed negative, where energy density was positive. Further, the condition $\rho+p_r$ has observed positive, where energy density was noticed negative. The condition $\rho+p_{r}+2p_{t}$ has also violated in some regions, which can be confirmed from Fig. (\ref{F9}) by left and right apart for both model-I and model-II, respectively. The presence of exotic matter, which is a crucial component for the existence of these proposed WH models, has confirmed by the violation of NEC and SEC. For both form function models, all of the energy conditions with positive as well as negative effects have provided in detail in the Tab. (\ref{tab1}).\\

The stability analysis for both shape functions through speed of sound parameters are investigated and it is found that both the considered shape functions are stable and viable Ricci-inverse gravitational frame. As a result, both WH solutions provide the stable configurations.

\section*{Appendix}
\begin{eqnarray*}
\rho _1&&=\frac{r (r-X(r))}{r \left(X'(r)-r a'(r)\right)+X(r) \left(r a'(r)+1\right)}+\frac{3 X(r)-3 r}{p_1},\\
\rho _2&&=\frac{r (r-X(r))}{r \left(2 r^2 a''(r)+r^2 a'(r)^2-r a'(r) X'(r)-4 X'(r)\right)+X(r) \left(-2 r^2 a''(r)+r^2 \left(-a'(r)^2\right)+r a'(r)+4\right)},\\
\rho _3&&=-2 r^4 (r-X(r))^2 (2 r-X(r)) X(r) a'(r)^6-r^3 (r-X(r)) a'(r)^5 \left(r^3 \left(5 X'(r)+4\right)+r^2 X(r) \left(15-26 X'(r)\right)\right.\\&&\left.+r X(r)^2 \left(13 X'(r)-10\right)-X(r)^3\right),\\
\rho _4&&=r^2 X(r)^2 \left(58 r^2 a''(r)-6 r X''(r)+13 X'(r)^2-44 X'(r)-170\right)+2 r X(r)^3 \left(-22 r^2 a''(r)+r X''(r)+3 X'(r)\right.\\&&\left.+67\right)+X(r)^4 \left(11 r^2 a''(r)-35\right)+r^4 \left(3 r^2 a''(r)+10 X'(r)^2-12 X'(r)-24\right)+2 r^3 X(r) \left(-14 r^2 a''(r)\right.\\&&\left.+2 r X''(r)-13 X'(r)^2+28 X'(r)+46\right),\\
\rho _5&&=-6 r^2 a^{(3)}(r)^2 (r-X(r))^2-a''(r)^2 \left(r^2 \left(4 r X''(r)+15 X'(r)^2-32 X'(r)+24\right)-2 r X(r) \left(2 r X''(r)-X'(r)+8\right)\right.\\&&\left.+7 X(r)^2\right)+3 r^2 (r-X(r))^2 a''(r)^3+2 r (r-X(r)) a''(r) \left(r a^{(4)}(r) (r-X(r))+a^{(3)}(r) \left(r \left(7 X'(r)-8\right)+X(r)\right)\right),
\end{eqnarray*}
\begin{eqnarray*}
\rho _6&&=r^4 \left(X'(r) \left(-64 r^2 a''(r)+26 r X''(r)+80\right)+4 \left(r^2 X^{(3)}(r)+3 r X''(r)-32\right)-3 X'(r)^3+72 X'(r)^2\right)\\&&-r^3 X(r) \left(X'(r) \left(-264 r^2 a''(r)+60 r X''(r)+352\right)+2 \left(8 r^3 a^{(3)}(r)+48 r^2 a''(r)+6 r^2 X^{(3)}(r)+15 r X''(r)\right.\right.\\&&\left.\left.-280\right)-3 X'(r)^3+151 X'(r)^2\right)+r^2 X(r)^2 \left(X'(r) \left(-300 r^2 a''(r)+30 r X''(r)+503\right)+4 \left(10 r^3 a^{(3)}(r)\right.\right.\\&&\left.\left.+34 r^2 a''(r)+3 r^2 X^{(3)}(r)+12 r X''(r)-202\right)+71 X'(r)^2\right)-r X(r)^3 \left(32 r^3 a^{(3)}(r)-5 \left(20 r^2 a''(r)-43\right) X'(r)\right.\\&&\left.+20 r^2 a''(r)+4 r^2 X^{(3)}(r)+26 r X''(r)-419\right)+X(r)^4 \left(8 r^3 a^{(3)}(r)-20 r^2 a''(r)-51\right),
\end{eqnarray*}
\begin{eqnarray*}
\rho _7&&=2 r^2 a''(r)^2 \left(r^3 \left(28-33 X'(r)\right)+r^2 X(r) \left(78 X'(r)-67\right)+X(r)^2 \left(30 r-39 r X'(r)\right)+3 X(r)^3\right)-4 r (r-X(r))\\&& \times\left(2 a^{(3)}(r) \left(r^2 \left(3 r X''(r)+X'(r)^2+20\right)-r X(r) \left(3 r X''(r)+2 X'(r)+40\right)+21 X(r)^2\right)-r a^{(4)}(r) (r-X(r))\right.\\&&\left.\times \left(r \left(X'(r)-4\right)+3 X(r)\right)\right)+a''(r) \left(\left(51-56 r^3 a^{(3)}(r)\right) X(r)^3+r^3 \left(4 \left(12 r^3 a^{(3)}(r)+r^2 X^{(3)}(r)-r X''(r)-48\right)\right.\right.\\&&\left.\left.-3 X'(r)^3+80 X'(r)^2+X'(r) \left(30 r X''(r)+64\right)\right)-r^2 X(r) \left(160 r^3 a^{(3)}(r)+8 r^2 X^{(3)}(r)+22 r X''(r)+71 X'(r)^2\right.\right.\\&&\left.\left.+6 X'(r) \left(5 r X''(r)+48\right)-512\right)+r X(r)^2 \left(168 r^3 a^{(3)}(r)+4 r^2 X^{(3)}(r)+26 r X''(r)+215 X'(r)-368\right)\right),\\
\rho _8&&=-24 r^4 a^{(4)}(r)-72 r^3 a^{(3)}(r)+204 r^4 a''(r)^2-6 r^2 a''(r) \left(10 r X''(r)+197\right)+\left(48 r^2 a''(r)+34\right) X'(r)^2-2 X'(r)\\&&\times \left(30 r^3 a^{(3)}(r)+96 r^2 a''(r)+r^2 X^{(3)}(r)-10 r X''(r)-173\right)+20 r^2 X^{(3)}(r)+6 r^2 X''(r)^2+259 r X''(r)+808,\\
\rho _9&&=16 r^4 a^{(4)}(r)+60 r^3 a^{(3)}(r)-144 r^4 a''(r)^2+4 r^2 a''(r) \left(5 r X''(r)+216\right)+2 \left(10 r^3 a^{(3)}(r)+16 r^2 a''(r)-69\right) \\&& \times X'(r)-6 r^2 X^{(3)}(r)-93 r X''(r)-654,\\
\rho _{10}&&=X'(r)^2 \left(-96 r^2 a''(r)+r X''(r)-74\right)+X'(r) \left(56 r^3 a^{(3)}(r)+260 r^2 a''(r)+4 r^2 X^{(3)}(r)-42 r X''(r)-272\right)\\&&-2 \left(-8 r^4 a^{(4)}(r)-24 r^3 a^{(3)}(r)+60 r^4 a''(r)^2-4 r^2 a''(r) \left(7 r X''(r)+87\right)+11 r^2 X^{(3)}(r)+6 r^2 X''(r)^2\right.\\&&\left.+119 r X''(r)+224\right)+2 X'(r)^3,\\
\rho _{11}&&=X'(r)^2 \left(42 r^2 a''(r)-r X''(r)+40\right)-2 X'(r) \left(8 r^3 a^{(3)}(r)+44 r^2 a''(r)+r^2 X^{(3)}(r)-11 r X''(r)-32\right)\\&&+2 \left(-2 r^4 a^{(4)}(r)-8 r^3 a^{(3)}(r)+12 r^4 a''(r)^2-8 r^2 a''(r) \left(r X''(r)+10\right)+4 r^2 X^{(3)}(r)+3 r^2 X''(r)^2\right.\\&&\left.+36 r X''(r)+48\right)-2 X'(r)^3,\;\;\;\;\;\;\rho _{12}=p_1^4 (r-X(r))^2,\\
\rho _{13}&&=2 \rho _4 r^2 a'(r)^4+\rho _6 r a'(r)^3+2 \rho _7 r (r-X(r)) a'(r)+2 a'(r)^2 \left(2 X(r)^4 \left(-2 r^4 a^{(4)}(r)-10 r^3 a^{(3)}(r)+18 r^4 a''(r)^2\right.\right.\\&&\left.\left.-112 r^2 a''(r)+99\right)+\rho _{11} r^4+\rho _{10} r^3 X(r)+\rho _8 r^2 X(r)^2+\rho _9 r X(r)^3\right)+\rho _3-8 \rho _5 r^2 (r-X(r))^2,
\end{eqnarray*}

\begin{eqnarray*}
p_1&&=2 r (r-X(r)) a''(r)-a'(r) \left(r \left(X'(r)-4\right)+3 X(r)\right)+r (r-X(r)) a'(r)^2,\\
p_2&&=(r-X(r))^2 a'(r) \left(r^2 (r-X(r)) a'(r)^3-r a'(r)^2 \left(r \left(3 X'(r)-4\right)+X(r)\right)+2 r \left(2 r a^{(3)}(r) (r-X(r))-a''(r) \right.\right.\\&&\left.\left.\times\left(3 r X'(r)+X(r)-4 r\right)\right)+2 a'(r) \left(r \left(3 r^2 a''(r)-r X''(r)-2 X'(r)-4\right)+X(r) \left(6-3 r^2 a''(r)\right)\right)\right),
\end{eqnarray*}

\begin{eqnarray*}
p_3&&=-3 r^3 a'(r)^5+r^2 a'(r)^4 \left(4 r^2 a''(r)+6 r+15\right)+4 \left(-12 r^5 a''(r)^3+(24-5 r) r^3 a''(r)^2+4 \left(6 r^4 a^{(4)}(r)\right.\right.\\&&\left.\left.+9 r^6 a^{(3)}(r)^2+74 r^3 a^{(3)}(r)+85\right)-2 r^2 \left(6 r^4 a^{(4)}(r)+20 r^3 a^{(3)}(r)-1\right) a''(r)\right)+r^2 a'(r)^3 \left(4 (r-6) r^2 a^{(3)}(r)\right.\\&&\left.-2 (7 r+58) r a''(r)-3 r+2\right)+r a'(r)^2 \left(-24 r^4 a^{(4)}(r)-4 (r+32) r^3 a^{(3)}(r)+8 (r+15) r^3 a''(r)^2+2 (5 r+51)\right.\\&&\left.\times r^2 a''(r)-17 r-256\right)+2 r a'(r) \left(12 r^4 a^{(4)}(r)+4 (17 r+12) r^2 a^{(3)}(r)-4 (r+8) r^3 a''(r)^2+r \left(4 (r+30)\right.\right.\\&&\left.\left.\times r^3 a^{(3)}(r)+27 r+520\right) a''(r)+88\right),
\end{eqnarray*}

\begin{eqnarray*}
p_4&&=-r^3 a'(r)^5 \left(X'(r)+11\right)+2 r^2 a'(r)^4 \left(8 r^2 a''(r)-r^2 X''(r)+3 r X''(r)+(3 r+10) X'(r)+9 r+20\right)\\&&-8 \left(24 r^5 a''(r)^3-9 r^3 a''(r)^2 \left(r^2 X''(r)-3 r+4\right)-4 \left(9 r^4 a^{(4)}(r)+18 r^6 a^{(3)}(r)^2-2 r \left(9 r^3 a^{(3)}(r)+31\right) X''(r)\right.\right.\\&&\left.\left.+126 r^3 a^{(3)}(r)-6 r^2 X^{(3)}(r)+119\right)+\left(-r^3 (17 r+12) a''(r)^2-4 \left(3 r^4 a^{(4)}(r)+22 r^3 a^{(3)}(r)+51\right)+r^2 \right.\right.\\&&\left.\left.\left(24 r^3 a^{(3)}(r)+127\right) a''(r)\right) X'(r)+r^2 a''(r) \left(24 r^4 a^{(4)}(r)+56 r^3 a^{(3)}(r)-12 r^2 X^{(3)}(r)+2 r X''(r)-131\right)\right)\\&&+r^2 a'(r)^3 \left(16 r^3 a^{(3)}(r)-96 r^2 a^{(3)}(r)-\left(2 r (5 r+18) a''(r)+5 (r-4)\right) X'(r)-2 (23 r+214) r a''(r)+12 r^2 X^{(3)}(r)\right.\\&&\left.+2 r^2 X''(r)+62 r X''(r)-7 r-12\right)-4 r a'(r)^2 \left(24 r^4 a^{(4)}(r)+3 r^4 a^{(3)}(r)+128 r^3 a^{(3)}(r)-8 (r+15) r^3 a''(r)^2\right.\\&&\left.+r^2 a''(r) \left((r+24) r X''(r)-7 r-68\right)+\left(r^4 a^{(3)}(r)-(3 r+34) r^2 a''(r)+16 r+88\right) X'(r)+3 r^3 X^{(3)}(r)\right.\\&&\left.-12 r^2 X^{(3)}(r)+15 r^2 X''(r)-46 r X''(r)+r+168\right)+2 r a'(r) \left(-4 (3 r+8) r^3 a''(r)^2-4 \left(-9 r^4 a^{(4)}(r)+6 r \right.\right.\\&&\left.\left.\times\left(3 r^3 a^{(3)}(r)+16\right) X''(r)-6 (11 r+6) r^2 a^{(3)}(r)+12 r^2 X^{(3)}(r)-95\right)+r a''(r) \left(16 (r+30) r^3 a^{(3)}(r)+12 r^3\right.\right.\\&&\left.\left.\times X^{(3)}(r)+2 (7 r-144) r X''(r)+127 r+1776\right)+\left(-4 (r+24) r^3 a''(r)^2-19 (r-16) r a''(r)+4 \left(3 r^4 a^{(4)}(r)\right.\right.\right.\\&&\left.\left.\left.+2 (r+6) r^2 a^{(3)}(r)-7\right)\right) X'(r)\right),
\end{eqnarray*}

\begin{eqnarray*}
p_5&&=-3 r^3 \left(X'(r)+5\right) a'(r)^5+3 r^2 \left(8 a''(r) r^2-2 X''(r) r^2+6 X''(r) r+6 r+3 X'(r)^2+2 (3 r+7) X'(r)+13\right)\\&&\times a'(r)^4+r^2 \left(-(r-42) X'(r)^2-\left(6 (5 r+18) a''(r) r-2 (r+21) X''(r) r+13 r+24\right) X'(r)+2 \left(12 a^{(3)}(r) r^3\right.\right.\\&&\left.\left.-72 a^{(3)}(r) r^2+18 X^{(3)}(r) r^2-3 (9 r+98) a''(r) r+2 (r+36) X''(r) r-2 r-3\right)\right) a'(r)^3-2 r \left(-24 (r+15)\right.\\&&\left. \times a''(r)^2 r^3+a''(r) \left(6 (r+24) X''(r) r-13 r-87\right) r^2+X'(r)^2 \left(37-(r-15) r a''(r)\right) r+2 X'(r) \left(3 a^{(3)}(r) r^4 \right.\right.\\&&\left.\left. +3 X^{(3)}(r) r^3-(8 r+117) a''(r) r^2+3 (3 r-14) X''(r) r+11 r+264\right)+2 \left(36 a^{(4)}(r) r^4-9 X''(r)^2 r^3+3 (r+64) \right.\right.\\&&\left.\left. \times a^{(3)}(r) r^3+6 X^{(3)}(r) r^3-36 X^{(3)}(r) r^2+12 (3 r-8) X''(r) r-4 r+120\right)\right) a'(r)^2+2 r \left(\left(-12 (r+24) a''(r)^2 r^3\right.\right.\\&&\left.\left. +a''(r) \left(66 X''(r) r^2-43 r+1344\right) r+4 \left(9 a^{(4)}(r) r^4+36 (r+1) a^{(3)}(r) r^2-12 X^{(3)}(r) r^2-72 X''(r) r+133\right)\right) \right.\\&&\left. X'(r)-X'(r)^2 \left(60 a^{(3)}(r) r^3+(7 r+216) a''(r) r+308\right)+4 \left(-54 X''(r) a^{(3)}(r) r^4+9 a^{(4)}(r) r^4-3 (r-8) \right.\right.\\&&\left.\left. \times a''(r)^2 r^3+81 a^{(3)}(r) r^3+36 X''(r)^2 r^2+36 a^{(3)}(r) r^2-24 X^{(3)}(r) r^2-216 X''(r) r+a''(r) \left(6 (r+30) a^{(3)}(r) r^3 \right.\right.\right.\\&&\left.\left. \left.+9 X^{(3)}(r) r^3-6 (r+36) X''(r) r+53 r+498\right) r+76\right)\right) a'(r)+4 \left(-72 a''(r)^3 r^5+3 a''(r)^2 \left(18 X''(r) r^2-31 r\right.\right.\\&&\left.\left.+24\right) r^3-4 a''(r) \left(18 a^{(4)}(r) r^4+24 a^{(3)}(r) r^3-18 X^{(3)}(r) r^2+36 X''(r) r-149\right) r^2+X'(r)^2 \left(39 a''(r)^2 r^4\right.\right.\\&&\left.\left. -178 a''(r) r^2-24 \left(5 a^{(3)}(r) r^3+8\right)\right)+8 \left(18 r^2 X''(r)^2-2 r \left(27 a^{(3)}(r) r^3+68\right) X''(r)+3 \left(9 a^{(3)}(r)^2 r^6\right.\right.\right.\\&&\left.\left.\left. +3 a^{(4)}(r) r^4+47 a^{(3)}(r) r^3-4 X^{(3)}(r) r^2+26\right)\right)+2 X'(r) \left(12 (r+3) a''(r)^2 r^3+a''(r) \left(-72 a^{(3)}(r) r^3\right.\right.\right.\\&&\left.\left.\left.+66 X''(r) r-203\right) r^2+4 \left(9 a^{(4)}(r) r^4+96 a^{(3)}(r) r^3-6 X^{(3)}(r) r^2-50 X''(r) r+201\right)\right)\right),
\end{eqnarray*}

\begin{eqnarray*}
p_6&&=-r^3 \left(X'(r)+2\right) a'(r)^5+r^2 \left(4 a''(r) r^2-2 X''(r) r^2+6 X''(r) r+9 X'(r)^2+(6 r+2) X'(r)+4\right) a'(r)^4\\&&+r^2 \left((r+24) X'(r)^3-2 (2 r+15) X'(r)^2+\left(-2 r (5 r+18) a''(r)+2 r (r+21) X''(r)+8\right) X'(r)+4 r \right.\\&&\left.\times \left(-(r+20) a''(r)+5 X''(r)+r \left((r-6) a^{(3)}(r)+3 X^{(3)}(r)\right)\right)\right) a'(r)^3+r \left(3 r X'(r)^4+(96-36 r) X'(r)^3\right.\\&&\left. +2 \left((r-15) a''(r) r^2+4 \left(3 X''(r) r^2+2 r-36\right)\right) X'(r)^2-4 \left(a^{(3)}(r) r^4+3 X^{(3)}(r) r^3-(2 r+49) a''(r) r^2\right.\right.\\&&\left.\left.+21 (r-2) X''(r) r+16\right) X'(r)+4 r \left(2 r^2 (r+15) a''(r)^2-r \left(r (r+24) X''(r)+16\right) a''(r)+9 r^2 X''(r)^2+4 X''(r)\right.\right.\\&&\left.\left.-2 r \left(3 a^{(4)}(r) r^2+16 a^{(3)}(r) r-6 X^{(3)}(r)\right)\right)\right) a'(r)^2+2 r \left(12 X'(r)^4+3 \left(13 r^2 a''(r)-76\right) X'(r)^3-4 \left(15 \right.\right.\\&&\left.\left.\times a^{(3)}(r) r^3+(31 r+54) a''(r) r-24 X''(r) r-76\right) X'(r)^2+2 r \left(-2 r^2 (r+24) a''(r)^2+\left(33 X''(r) r^2+56 r+368\right)\right.\right.\\&&\left. \left. \times a''(r)-240 X''(r)+2 r \left(4 (8 r+3) a^{(3)}(r)+3 \left(r^2 a^{(4)}(r)-4 X^{(3)}(r)\right)\right)\right) X'(r)+4 r^2 \left(16 r a''(r)^2+\left(r^2 \right.\right.\right.\\&&\left.\left.\left. \times \left((r+30) a^{(3)}(r)+3 X^{(3)}(r)\right)-(13 r+72) X''(r)\right) a''(r)+18 X''(r) \left(2 X''(r)-r^2 a^{(3)}(r)\right)\right)\right) a'(r)+4 \\&&\times\left(12 X'(r)^4+\left(78 r^2 a''(r)-296\right) X'(r)^3+\left(39 a''(r)^2 r^4-412 a''(r) r^2+24 \left(-5 a^{(3)}(r) r^3+4 X''(r) r+26\right)\right) \right.\\&& \left.\times X'(r)^2-4 r \left(r^2 (11 r-6) a''(r)^2+3 r \left(4 a^{(3)}(r) r^3-11 X''(r) r-28\right) a''(r)+2 \left(-3 a^{(4)}(r) r^3-52 a^{(3)}(r) r^2 \right.\right.\right.\\&&\left.\left.\left. +6 X^{(3)}(r) r+74 X''(r)\right)\right) X'(r)-2 r^2 \left(6 a''(r)^3 r^3-9 a''(r)^2 X''(r) r^3+2 a''(r) \left(3 a^{(4)}(r) r^3-2 a^{(3)}(r) r^2\right.\right.\right.\\&&\left.\left.\left.-6 X^{(3)}(r) r+34 X''(r)\right) r-18 \left(r^2 a^{(3)}(r)-2 X''(r)\right)^2\right)\right),\\
p_7&&=-2 (r-40) r^2 a''(r)^2+2 \left(-6 \left(9 r^3 a^{(3)}(r)+20\right) X''(r)+r \left(3 \left(r^2 a^{(4)}(r)-4 X^{(3)}(r)\right)+4 (8 r+3) a^{(3)}(r)\right) \right.\\&&\left.+72 r X''(r)^2\right)+a''(r) \left(8 (r+30) r^3 a^{(3)}(r)+18 r^3 X^{(3)}(r)-9 (5 r+48) r X''(r)+56 r+368\right),\\
p_8&&=3 \left(13 r^2 a''(r)-60\right) X'(r)^3-X'(r)^2 \left(120 r^3 a^{(3)}(r)+(131 r+432) r a''(r)-96 r X''(r)+76\right)-4 X'(r) \\&& \times \left(-9 r^4 a^{(4)}(r)-66 r^3 a^{(3)}(r)-36 r^2 a^{(3)}(r)+3 (r+24) r^3 a''(r)^2-r a''(r) \left(33 r^2 X''(r)+22 r+444\right)\right.\\&&\left.+24 r^2 X^{(3)}(r)+192 r X''(r)-152\right)+2 p_7 r,
\end{eqnarray*}

\begin{eqnarray*}
p_9&&=2 p_8 r a'(r)-3 r^3 a'(r)^5 \left(X'(r)+3\right)+2 r^2 a'(r)^4 \left(8 r^2 a''(r)-3 r^2 X''(r)+9 r X''(r)+9 X'(r)^2+3 (3 r+4) X'(r)\right.\\&&\left.+3 r+9\right)+8 \left(\left(39 r^2 a''(r)-124\right) X'(r)^3+X'(r)^2 \left(-120 r^3 a^{(3)}(r)+39 r^4 a''(r)^2-295 r^2 a''(r)+48 r X''(r)+180\right)\right.\\&&\left.+r \left(-24 r^4 a''(r)^3+r^2 a''(r)^2 \left(27 r^2 X''(r)-22 r+12\right)+4 \left(-2 \left(27 r^3 a^{(3)}(r)+37\right) X''(r)+r \left(3 r^2 a^{(4)}(r)\right.\right.\right.\right.\\&&\left.\left.\left.\left.+18 r^4 a^{(3)}(r)^2+52 r a^{(3)}(r)-6 X^{(3)}(r)\right)+36 r X''(r)^2\right)-2 r a''(r) \left(12 r^4 a^{(4)}(r)+4 r^3 a^{(3)}(r)-18 r^2 X^{(3)}(r)\right.\right.\right.\\&&\left.\left.\left.+69 r X''(r)-84\right)\right)+X'(r) \left(36 r^4 a^{(4)}(r)+504 r^3 a^{(3)}(r)-9 (3 r-4) r^3 a''(r)^2+4 r^2 a''(r) \left(-18 r^3 a^{(3)}(r)+33 r \right.\right.\right.\\&&\left.\left.\left. \times X''(r)+23\right)-48 r^2 X^{(3)}(r)-496 r X''(r)+624\right)\right)+r^2 a'(r)^3 \left(-2 X'(r) \left(3 (5 r+18) r a''(r)-2 (r+21) r X''(r) \right.\right.\\&&\left.\left. +4 r+18\right)+2 \left(8 r^3 a^{(3)}(r)-48 r^2 a^{(3)}(r)-(13 r+178) r a''(r)+18 r^2 X^{(3)}(r)+(r+51) r X''(r)+4\right)+(r+24)\right.\\&&\left. \times X'(r)^3+(12-5 r) X'(r)^2\right)-4 r a'(r)^2 \left(24 r^4 a^{(4)}(r)+r^4 a^{(3)}(r)+128 r^3 a^{(3)}(r)-8 (r+15) r^3 a''(r)^2+r^2 a''(r) \right.\\&&\left. \times \left(3 (r+24) r X''(r)-2 r+15\right)+X'(r)^2 \left(-(r-15) r^2 a''(r)-6 r^2 X''(r)+19 r+72\right)+X'(r) \left(3 r^4 a^{(3)}(r)\right.\right.\\&&\left.\left.-(7 r+132) r^2 a''(r)+6 r^3 X^{(3)}(r)+6 (5 r-14) r X''(r)-8 r+192\right)+3 r^3 X^{(3)}(r)-18 r^3 X''(r)^2-36 r^2 X^{(3)}(r)\right.\\&&\left.+21 r^2 X''(r)-54 r X''(r)+6 (r-4) X'(r)^3+16\right),
\end{eqnarray*}
\begin{eqnarray*}
p_{10}&&=r \left(-2 r^2 a''(r)+r a'(r) X'(r)-r a'(r)^2+4 X'(r)\right)+X(r) \left(2 r^2 a''(r)+r a'(r)^2-r a'(r)-4\right),\\
p_{11}&&=\frac{(r-X(r))^2 \left(2 r a''(r)-\frac{8 \alpha  r^5}{(r-X(r)) \left(r \left(X'(r)-r a'(r)\right)+X(r) \left(r a'(r)+1\right)\right)}+\frac{\left(r a'(r)+2\right) \left(r^2 a'(r)-r X(r) a'(r)-r X'(r)+X(r)\right)}{r (r-X(r))}\right)}{r^3},\\
p_{12}&&=\frac{r-X(r)}{r^2 \left(-a'(r)\right)+r X(r) a'(r)+r X'(r)+X(r)},
\end{eqnarray*}
\begin{eqnarray*}
p_{13}&&=\frac{\frac{X(r)-r}{p_1}-\frac{r (r-X(r))}{r \left(2 r^2 a''(r)+r^2 a'(r)^2-r a'(r) X'(r)-4 X'(r)\right)+X(r) \left(-2 r^2 a''(r)+r^2 \left(-a'(r)^2\right)+r a'(r)+4\right)}}{r},\\
p_{14}&&=r^2 (r-X(r)) a'(r)^3-r a'(r)^2 \left(r \left(3 X'(r)-4\right)+X(r)\right)+2 r \left(2 r a^{(3)}(r) (r-X(r))-a''(r) \left(3 r X'(r)+X(r)\right.\right.\\&&\left.\left.-4 r\right)\right)+2 a'(r) \left(r \left(3 r^2 a''(r)-r X''(r)-2 X'(r)-4\right)+X(r) \left(6-3 r^2 a''(r)\right)\right),\\
p_{15}&&=-6 r^3 a^{(3)}(r)+36 r^4 a''(r)^2-36 r^3 a''(r) X''(r)+\left(4-4 r^2 a''(r)\right) X'(r)^2-71 r^2 a''(r)-r^3 a'(r)^3 \left(X'(r)+2\right)\\&&+X'(r) \left(6 r^3 a^{(3)}(r)+8 r^2 a''(r)-2 r^2 X^{(3)}(r)-2 r X''(r)+8\right)+r^2 a'(r)^2 \left(3 r^2 a''(r)+5 r X''(r)+6 X'(r)^2\right.\\&&\left.+18 X'(r)-8\right)+r a'(r) \left(X'(r) \left(-24 r^2 a''(r)+10 r X''(r)+35\right)-2 \left(6 r^3 a^{(3)}(r)+16 r^2 a''(r)-3 r^2 X^{(3)}(r)\right.\right.\\&&\left.\left.-14 r X''(r)+6\right)\right)+4 r^2 X^{(3)}(r)+6 r^2 X''(r)^2+47 r X''(r)+12,
\end{eqnarray*}
\begin{eqnarray*}
p_{16}&&=r^3 X(r) \left(\left(10-9 r^2 a''(r)\right) X'(r)^2-r^3 a'(r)^3 \left(2 X'(r)+1\right)+2 r \left(r \left(X^{(3)}(r)-r a^{(3)}(r)\right)+12 r^3 a''(r)^2-6 r a''(r)\right.\right.\\&&\left.\left.\times \left(3 r X''(r)+2\right)+6 r X''(r)^2+11 X''(r)\right)+X'(r) \left(6 r^3 a^{(3)}(r)+4 r^2 a''(r)-4 r^2 X^{(3)}(r)+8\right)+r^2 a'(r)^2 \left(3 r^2 a''(r)\right.\right.\\&&\left.\left.+4 r X''(r)+12 X'(r)^2+2 X'(r)-6\right)+r a'(r) \left(-4 X'(r) \left(6 r^2 a''(r)-5 r X''(r)-3\right)-2 \left(4 r^3 a^{(3)}(r)+5 r^2 a''(r)\right.\right.\right.\\&&\left.\left.\left.-3 r^2 X^{(3)}(r)-5 r X''(r)+6\right)+5 X'(r)^2\right)\right)+r^4 \left(-\left(2 r X'(r) \left(r^2 a^{(3)}(r)-r X^{(3)}(r)+X''(r)\right)+6 r^2 \left(X''(r)\right.\right.\right.\\&&\left.\left.\left.-r a''(r)\right)^2+X'(r)^2 \left(-5 r^2 a''(r)+r X''(r)+4\right)-r^3 a'(r)^3 X'(r)+r^2 a'(r)^2 \left(r \left(r a''(r)+X''(r)\right)+6 X'(r)^2\right.\right.\right.\\&&\left.\left.\left.-4 X'(r)\right)+r a'(r) \left(-2 r \left(r^2 a^{(3)}(r)-r X^{(3)}(r)+X''(r)\right)-2 X'(r) \left(4 r^2 a''(r)-5 r X''(r)+2\right)+X'(r)^3\right.\right.\right.\\&&\left.\left.\left.+4 X'(r)^2\right)\right)\right)+r X(r)^3 \left(2 \left(r^3 a^{(3)}(r) \left(X'(r)-3\right)+r^2 X^{(3)}(r)+13 r X''(r)+15\right)+24 r^4 a''(r)^2+r^2 a''(r) \right.\\&&\left.\times\left(-12 r X''(r)+4 X'(r)-71\right)-r^3 a'(r)^3+r^2 a'(r)^2 \left(r^2 a''(r)+2 r X''(r)+12 X'(r)+4\right)+r a'(r) \left(-8 r^3 a^{(3)}(r)\right.\right.\\&&\left.\left.+\left(20-8 r^2 a''(r)\right) X'(r)-34 r^2 a''(r)+2 r^2 X^{(3)}(r)+16 r X''(r)+19\right)\right)+2 X(r)^4 \left(r^3 a^{(3)}(r)-3 r^4 a''(r)^2\right.\\&&\left.+12 r^2 a''(r)-3 r^2 a'(r)^2+r a'(r) \left(r^3 a^{(3)}(r)+6 r^2 a''(r)-10\right)-10\right)-p_{15} r^2 X(r)^2,\\
p_{17}&&=\frac{p_{16}}{\left(r \left(X'(r)-r a'(r)\right)+X(r) \left(r a'(r)+1\right)\right)^4}+\frac{2 p_{14} (r-X(r))^2 a'(r)}{p_1^3 r}.
\end{eqnarray*}

\end{document}